\documentclass[onecolumn,showpacs]{revtex4}

\usepackage{graphicx}
\usepackage{dcolumn}
\usepackage{bm}

\input epsf

\begin{document}

\title{\Large Naked Singularities in Higher Dimensional Gravitational Collapse}

\author{\bf Asit Banerjee}
\email{asitb@cal3.vsnl.net.in}
\affiliation{Department of Physics, Jadavpur University,
Calcutta-32, India.}
\author{\bf Ujjal Debnath}
\email{ujjaldebnath@yahoo.com}
\author{\bf Subenoy Chakraborty}
\email{subenoyc@yahoo.co.in}
\affiliation{Department of
Mathematics, Jadavpur University, Calcutta-32, India.}

\date{\today}

\begin{abstract}
Spherically symmetric inhomogeneous dust collapse has been studied
in higher dimensional space-time and the factors responsible for
the appearance of a naked singularity are analyzed in the region
close to the centre for the marginally bound case. It is clearly
demonstrated that in the former case naked singularities do not
appear in the space-time having more than five dimension, which
appears to a strong result. The non-marginally bound collapse is
also examined in five dimensions and the role of shear in
developing naked singularities in this space-time is discussed in
details. The five dimensional space-time is chosen in the later
case because we have exact solution in closed form only in five
dimension and not in any other case.
\end{abstract}


\maketitle

\section{\normalsize\bf{Introduction}}
The Cosmic Censorship Conjecture [1,2] is yet one of the
unresolved problems in General Relativity. It states that the
space-time singularity produced by gravitational collapse must be
covered by the horizon. However the singularity theorems as such
do not state anything about the visibility of the singularity to
an outside observer. In fact the conjecture has not yet any
precise mathematical proof. Several models related to the
gravitational collapse of matter has so far been constructed
where one encounters a naked singularity [3-8].\\

It has recently been pointed out by Joshi et al [9] that the
physical feature which is responsible for the formation of naked
singularity is nothing but the presence of shear. It is the shear
developing in the gravitational collapse, which delays the
formation of the apparent horizon so that the communication is
possible from the very strong gravity region to observers
situated outside. Joshi et al have analyzed in details how the
presence of shear determines the growth and evolution of an
inhomogeneous dust distribution represented by four dimensional
Tolman-Bondi metric and have attempted to clarify the nature of
singularities as the final outcome.\\

The objective of this paper is to fully investigate the situation
in the background of higher dimensional space-time with arbitrary
dimensions for marginally bound collapse and in five dimensional
space-time for non-marginally bound case. The reason for
confining our discussions to five dimensions only in the second
case is that the closed form solutions for non-marginally bound
collapse are available only in $5D$ space-time. These discussions
are of adequate relevance in the context of recent interest
generated in the study of gravitational collapse in higher
dimensional space-time [10-16].\\

We have shown in the present paper that for marginally bound
collapse that is for $f=0$ the naked singularity may appear only
when the space-time has dimensions upto five which appears to a
strong result. In more than five dimensions $t_{ah}<t_{0}$ that is
the apparent horizon at any point in the region forms earlier than
the central shell focusing singularity, which indicates that the
shell focusing singularity first appearing at $r=0$ remains hidden
behind the
apparent horizon and so gives rise to a black hole always.\\

The non-marginally bound collapse of a $5D$ inhomogeneous dust
reveals that there may occur both the black hole and the naked
singularities depending on the initial data. The five dimensional
space-time is chosen in the later case because we have exact
solution in closed form only in five dimension and not in any
other case.

\section{\normalsize\bf{Higher dimensional inhomogeneous Dust}}
 The higher dimensional Tolman-Bondi type metric is given by

\begin{equation}
ds^{2}=e^{\nu}dt^{2}-e^{\lambda}dr^{2}-R^{2}d\Omega^{2}_{n}
\end{equation}

where $\nu,\lambda,R$ are functions of the radial co-ordinate $r$
and time $t$ and $d\Omega^{2}_{n}$ represents the metric on the
$n$-sphere. Since we assume the matter in the form of dust, the
motion of particles will be geodesic allowing us to write

\begin{equation}
e^{\nu}=1
\end{equation}

Using comoving co-ordinates one can in view of the field
equations [11] arrive at the following relations in ($n+2$)
dimensional space-time

\begin{equation}
e^{\lambda}=\frac{R'^{2}}{1+f(r)}
\end{equation}

and

\begin{equation}
\dot{R}^{2}=f(r)+\frac{F(r)}{R^{n-1}},
\end{equation}

where the function $f(r)$ classifies the space-time [2] as bound,
marginally bound and unbound depending on the range of its values
which are respectively

\begin{equation}
f(r)<0,~~~f(r)=0,~~~f(r)>0.
\end{equation}

The function $F(r)$ can be interpreted as the mass function which
is related with the mass contained within the comoving radius $r$.

The above models are characterized by the initial data specified
on the initial hypersurface $t=t_{i}$, from which the collapse
develops. As it is possible to make an arbitrary relabeling of
spherical dust shells by $r\rightarrow g(r)$, without loss of
generality we fix the labeling by the choice $R(t_{i},r)=r$, so
that initial density distribution is given by

\begin{equation}
\rho_{i}(r)=\frac{n F'}{2r^{n}}
\end{equation}

Now the integration of the equation (4) gives us the exact
solution for $R$ as a function of $r$ and $t$. For the case $f=0$
the integration is straightforward to yield the solution

\begin{equation}
R=\left(\frac{n+1}{2}\right)^{\frac{2}{n+1}}F^{\frac{1}{n+1}}(t_{s}-t)^{\frac{2}{n+1}}
\end{equation}

where $t_{s}$ is a function of the radial co-ordinate $r$ and
$t=t_{s}$ is the instant of shell focusing singularity occurring
at $r$, that is $R(t_{s}(r),r)=0$. It indicates that the shell
focusing singularity occurs at different $r$ at different epochs.
So the collapse in this case is not simultaneous in comoving
co-ordinates and is in variance with that in the homogeneous dust
model [6,7,12]. The second case is for $f\neq 0$. One can not,
however, obtain the explicit solution of (4) in closed form except
for $n=3$ that is in five dimensional space-time. Even in the
simplest case of 4 dimensions the solution for non-zero $f(r)$ is
obtained in the parametric form only.

\section{\normalsize\bf{Marginally bound case (}$f=0$\bf{) in (}$n+2$\bf{)}$D$ \bf{space-time}}

We observe that the normal vector to the boundary of any
hypersphere $R-R_{0}=0$ is given by
$l_{\mu}=(\dot{R},R',0,0,0,......)$. In order that $l_{\mu}$ is a
null vector and the boundary of this hypersphere is a null surface
we must have

\begin{equation}
R(t_{ah},r)=[F(r)]^{\frac{1}{n-1}}
\end{equation}

where $t=t_{ah}(r)$ is the equation of apparent horizon which
marks the boundary of the trapped region. If the apparent horizon
develops earlier than the time of singularity formation, the
event horizon can fully cover the singularity, which then may be
said to be hidden within a black hole. Combining (7) and (8) along
with the initial condition $R(t_{i},r)=r$ we arrive at the
relation

\begin{equation}
(t_{ah}-t_{i})=\frac{2}{n+1}r^{\frac{n+1}{2}}F^{^{-1/2}}-\frac{2}{n+1}F^{\frac{1}{n-1}}
\end{equation}

Since our study is restricted to the region near $r=0$, the mass
function $F(r)$ should vanishes exactly at $r=0$ and it is
possible to express $F(r)$ as a polynomial function of $r$ near
the origin. One should at the same time keep in mind that
initially the central density is regular. Further from physical
considerations one may argue that $\rho_{i}'(r)$ should vanish
exactly at $r=0$, but is negative in the neighbouring region. All
these consideration lead to the following expressions for $F(r)$
and the initial energy density $\rho_{i}$,

\begin{equation}
F(r)=\left[F_{0}r^{n+1}+F_{2}r^{n+3}+F_{3}r^{n+4}.........\right]
\end{equation}

and

\begin{equation}
\rho_{i}(r)=\frac{n}{2}\left[(n+1)F_{0}+(n+3)F_{2}r^{2}+.........\right]
\end{equation}

Obviously the initial central density is given by
$\rho_{c}=\frac{n(n+1)}{2}F_{0}$. Since $\rho_{i}'(r)<0$ in the
region $r\approx 0$ one must have $F_{2}<0$.\\

Now in view of (7) one can write

\begin{equation}
t_{s}-t_{i}=\frac{2}{(n+1)}\frac{r^{(n+1)/2}}{\left[F_{0}r^{n+1}+F_{2}r^{n+3}+...\right]^{1/2}}
\end{equation}

If we now denote $t_{s}=t_{0}$ as the instant of first shell
focusing singularity to occur at the centre $r=0$, the relation
(12) can be used to obtain the following relation

\begin{equation}
t_{0}-t_{i}=\frac{2}{(n+1)F_{0}^{1/2}},
\end{equation}

which in combination with the relation (9) yields

\begin{eqnarray*}
t_{ah}-t_{0}=-\left[\frac{1}{n+1}\frac{F_{2}}{F_{0}^{3/2}}r^{2}+
\frac{1}{n+1}\frac{F_{3}}{F_{0}^{3/2}}r^{3}+\frac{1}{n+1}\frac{F_{4}}{F_{0}^{3/2}}r^{4}+...\right]
\end{eqnarray*}
\vspace{-5mm}

\begin{equation}
-\left[\frac{2}{n+1}F_{0}^{\frac{1}{n-1}}r^{\frac{n+1}{n-1}}+
\frac{2}{n^{2}-1}\frac{F_{2}}{F_{0}^{\frac{n-2}{n-1}}}r^{\frac{3n-1}{n-1}}+
\frac{2}{n^{2}-1}\frac{F_{3}}{F_{0}^{\frac{n-2}{n-1}}}r^{\frac{4n-2}{n-1}}+...\right]
\end{equation}

The relation (14) is valid in general $(n+2)$ dimensional
space-time and reduces to the 4 dimensional expression $(n=2)$ in
the region near the centre as given in the paper of Joshi et al
[9]. In $5D$ space-time the naked singularity appears only if
$F_{2}\ne 0$ and further $|F_{2}|>2F_{0}^{2}$. Otherwise for
either $F_{2}=0$ or $|F_{2}|<2F_{0}^{2}$ there must occur a black
hole, because in that case $t_{ah}<t_{0}$ that is, the horizon
appears earlier than the shell focusing singularity at $r=0$.
Only in a very special case of $|F_{2}|=2F_{0}^{2}$ the occurrence
of naked singularity or black hole will depend on the
co-efficients of higher powers of $r$. For general $(n+2)$
dimensions the close examination of (14) reveals that if there is
to exist a naked singularity we must have

\begin{equation}
\frac{n+1}{n-1}\geq 2,
\end{equation}

which means $n\leq 3$. It is interesting to note that $\left(
\frac{n+1}{n-1}\right)$ has the value 2 when $n=3$ and then it
decreases monotonically with the increasing number of dimensions.
So for space-time with dimension larger than five the term
$\frac{2}{n+1}F_{0}^{\frac{1}{n-1}}r^{\frac{n+1}{n-1}}$ dominates
leading to a black hole. Our conclusion is that for the space-time
having larger than five dimensions the existence of the naked
singularity is prohibited or in other words the shell focusing
singularity is fully covered by the horizon in such cases.
However, if we relax the restriction namely, $\rho'_{i}(r)=0$ at
$r=0$ then $F_{1}$ will be non-zero and this will leads to the
possibility of naked singularity in any dimension.\\

{\bf Calculation of shear near $r=0$:}\\

We still consider the marginally bound case $f=0$. The shear
scalar in the $(n+2)$ dimensional spherically symmetric dust
metric so far discussed in this section may be estimated by the
factor [9]
$\sigma=\sqrt{\frac{n}{2(n+1)}}\left(\frac{\dot{\lambda}}{2}-
\frac{\dot{R}}{R}\right)$. Using (3) and (4) one gets in turn

\begin{equation}
\sigma=\sqrt{\frac{n}{2(n+1)}}\left(\frac{\dot{R}'}{R'}-\frac{\dot{R}}{R}\right)
=\sqrt{\frac{n}{8(n+1)}}\frac{[R
F'-(n+1)R'F]}{F^{1/2}R^{(n+1)/2}R'}
\end{equation}

From (7)

$$
R'=\left[\frac{n+1}{2}\right]^{\frac{2}{n+1}}\left[\left(F^{\frac{1}{n+1}}\right)'(t_{s}-t)^{\frac{2}{n+1}}+\frac{2}{n+1}F^{\frac{1}{n+1}}(t_{s}-t)^{\frac{2}{n+1}-1}t_{s}'\right]
$$

where $F, F', t_{s}$ etc.'s are all functions of the comoving
radial co-ordinate $r$. So as $t$ approaches $t_{s}, R$
approaches zero and effectively $\sigma^{2}$ approaches
$(n+1)^{2}\frac{F}{R^{n+1}}$ , which in turn becomes infinitely
large. It is therefore evident that at each $r$, the magnitude of
shear explodes as the energy density explodes ($R=0$) and this
occurs at different instants at different spherical shells of
different radial co-ordinates including the centre ($r=0$).\\

It is now possible in view of the choice $R(t_{i},r)=r$ to
express the initial shear $\sigma_{i}$ to the form

\begin{equation}
\sigma_{i}=\sqrt{\frac{n}{8(n+1)}}\frac{[r F'-(n+1)F
]}{F^{1/2}r^{(n+1)/2}}
\end{equation}

But using (7) along with the initial condition $R(t_{i},r)=r$, we
have a relation like

\begin{equation}
t-t_{i}=F^{^{-1/2}}\left[r^{\frac{n+1}{2}}-R^{\frac{n+1}{2}}\right]
\end{equation}

which in turn being used in (16) yields the expression for shear
scalar $\sigma(t,r)$ very near the centre $(r\approx 0)$ in the
following form after some manipulations:

\begin{equation}
\sigma=\sqrt{\frac{n}{8(n+1)}}\frac{\sum_{m=2}^{\infty}m
F_{m}r^{m}}{F_{0}^{1/2}\left[1+\frac{(n+1)^{2}}{4}F_{0}(t-t_{i})^{2}
-(n+1)F_{0}^{1/2}(t-t_{i})\right]}
\end{equation}

so the initial shear at $t=t_{i}$ is given by

\begin{equation}
\sigma_{i}=\sqrt{\frac{n}{8(n+1)}}\frac{\sum_{m=2}^{\infty}m
F_{m}r^{m}}{F_{0}^{1/2}}
\end{equation}

We find that at the centre $(r=0)$ the initial shear vanishes. It
is interesting to note that  the dependence on $r$ of the initial
shear does not depend on the number of dimensions of the
space-time. In fact the expression (20) coincides exactly with the
value of the initial shear $\sigma_{i}$ calculated earlier by
Joshi et al [9] for $4D$ space-time $(n=2)$. In view of what has
been discussed so far the expression (20)  reveals that the
existence of the naked singularity is directly related with the
non-vanishing shear in four and five dimensions. But in larger
dimensions the central singularity seems to be covered  by the
appearance of apparent horizons. One should mention here that the
statement of the paper of Joshi et al [9] that the shear
decreases in course of time and finally vanishes at $t=t_{0}$ is
not true. In fact the shear in view of (19) increases and goes to
infinity at this epoch, which is expected.

\section{\normalsize\bf{Non-marginally bound five dimensional space-time} $(f\neq 0)$}

We already know that for $f\neq 0$ it is possible to obtain the
solution of (4) in closed form only in $5D$ space-time that is
for $n=3$. In other cases even for the simplest case of a four
dimensional manifold the solution is available only in parametric
form. In $5D$ space-time one of the exact solution of the
equation (4) is given by

\begin{eqnarray}
R^{2}=\left[f(t_{s}-t)^{2}+2F^{1/2}(t_{s}-t)\right]
\end{eqnarray}

For $f=0$ the solution (21) is coincident with the solution (7)
when we put $n=3$. By the same arguments put forward in the
previous section it is possible to express $F(r)$ as

\begin{equation}
F(r)=F_{0}r^{4}+F_{2}r^{6}+F_{3}r^{7}+.........
\end{equation}

and the initial density distribution $\rho_{i}(r)$ as

\begin{equation}
\rho_{i}(r)=\rho_{c}+\rho_{2}r^{2}+\rho_{3}r^{3}+.........,
\end{equation}

where and $\rho_{c}=6F_{0}$, $\rho_{2}=9F_{2}$, so on.\\

Here also following the earlier reasoning $F_{2}<0$. The only new
input in the present case is the function $f(r)$, which may be
expressed as a power series in $r$ near the centre $r=0$.\\

We assume

\begin{equation}
f(r)=f_{0}r+f_{1}r^{2}+f_{2}r^{3}+.........
\end{equation}

The form (24) is chosen because $f(r)$ vanishes as $r\rightarrow
0$, which is demanded by the regularity condition at $r=0$ [17].\\

Now when $t=t_{i}$ one can write from (21)

\begin{equation}
t_{s}-t_{i}=\frac{(F+f r^{2})^{1/2}}{f}-\frac{F^{1/2}}{f}
\end{equation}

The finite  non zero magnitude of the left hand side of (25)
demands $f_{0}=0$ and $f_{1}\neq 0$ as minimum requirements. The
expansion of (22) near $r=0$ yields

\begin{eqnarray*}
t_{s}-t_{i}=\frac{(F_{0}+f_{1})^{1/2}-F_{0}^{1/2}}{f_{1}}+\frac{f_{2}}{f_{1}^{2}}\left[\frac{2F_{0}^{1/2}(F_{0}+f_{1})^{1/2}-
(2F_{0}+f_{1})}{2(F_{0}+f_{1})^{1/2}}\right]r+\left[-\frac{f_{2}^{2}}{2f_{1}^{2}\sqrt{f_{1}+F_{0}}}\right.
\end{eqnarray*}
\vspace{-5mm}

\begin{equation}
\left.+\frac{(f_{2}^{2}-f_{1}f_{3})
(\sqrt{f_{1}+F_{0}}-\sqrt{F_{0}})}{f_{1}^{3}}-\frac{F_{2}}{2f_{1}\sqrt{F_{0}}}+
\frac{4(f_{1}+F_{0})(f_{3}+F_{2})-f_{2}^{2}}{8f_{1}(f_{1}+F_{0})^{3/2}}\right]r^{2}+.........
\end{equation}

At the centre $r=0$ the relation (26) reduces exactly to

\begin{equation}
t_{0}-t_{i}=\frac{(F_{0}+f_{1})^{1/2}-F_{0}^{1/2}}{f_{1}}
\end{equation}

Now at $t=t_{ah}$ one must have $R^{2}=F$ as justified earlier.
Hence by solving (21) we get

\begin{equation}
t_{ah}(r)-t_{s}(r)=\frac{F^{1/2}-F^{1/2}(1+f)^{1/2}}{f}
\end{equation}

The form (28) is chosen because it satisfies the consistency
relation in the limit $f\approx 0$. Remembering that
$t_{s}(r=0)=t_{0}$ we obtain in the region very close to the
centre

\begin{eqnarray*}
t_{ah}(r)=t_{0}+\frac{f_{2}}{f_{1}^{2}}\left[\frac{2F_{0}^{1/2}(F_{0}+f_{1})^{1/2}-
(2F_{0}+f_{1})}{2(F_{0}+f_{1})^{1/2}}\right]r+\left[-\frac{f_{2}^{2}}{2f_{1}^{2}
\sqrt{f_{1}+F_{0}}}-\frac{f_{1}F_{0}+F_{2}}{2f_{1}\sqrt{F_{0}}}\right.
\end{eqnarray*}
\vspace{-5mm}

\begin{equation}
\left.+\frac{(f_{2}^{2}-f_{1}f_{3})
(\sqrt{f_{1}+F_{0}}-\sqrt{F_{0}})}{f_{1}^{3}}+
\frac{4(f_{1}+F_{0})(f_{3}+F_{2})-f_{2}^{2}}{8f_{1}(f_{1}+F_{0})^{3/2}}\right]r^{2}+.........
\end{equation}

In the above relation if $f_{2}<0$ the second term on the R.H.S
is positive, which can easily be verified. Hence it is possible
to conclude that $t_{ah}>t_{0}$ near $r=0$. This suggests that
that the apparent horizon in the region close to the centre
appears later than the shell focusing singularity at $r=0$ and
hence the naked singularity should exist in this case. On the
other hand if $f_{2}>0$ it should be a black hole. These
conclusions are valid irrespective of $f(r)$ be positive or
negative except for the restriction $F_{0}>|f_{1}|$ in case
$f_{1}<0$. However, no definite conclusion can be drawn in case
$f_{2}=0$.\\

{\bf Calculation of shear:}\\

The general expression for the shear in five dimensional
space-time is given by

\begin{equation}
\sigma=\sqrt{\frac{3}{32}}\left(f+\frac{F}{R^{2}}\right)^{-1/2}\left[\frac{f'}{R'}+
\frac{F'}{R^{2}R'}-\frac{4F}{R^{3}}-\frac{2f}{R}\right]
\end{equation}

Using the expansions of $F(r)$ and $f(r)$ from (22) and (24) along
with the assumption $f_{0}=0$, which has already been justified
earlier we arrive at the expression for the shear at any instant
in the region close to $r=0$. This expression is given after
omitting a few intermediate steps by

\begin{equation}
\sigma=\sqrt{\frac{3}{32}}(F_{0}+f_{1})^{^{-1/2}}\left[\frac{f_{2}}{X}~r+
m\left(\frac{F_{m}}{X^{3}}+\frac{f_{m+1}}{X}\right)r^{m}+O(r^{m+1})\right]
\end{equation}

where $m\geq 2$ and
$X=X(t)=[1+f_{1}(t-t_{i})^{2}-2(F_{0}+f_{1})^{1/2}(t-t_{i})]$.\\

Since $\sigma$ depends on the time the shear develops in course of
time at any $r$ away from the centre. When $f=0$ we get for the
initial shear at $t=t_{i}$
$$
\sigma=\sqrt{\frac{3}{32}}~\frac{m
F_{m}}{F_{0}^{1/2}}r^{m}+O(r^{m+1})
$$
with $m\geq 2$. This is the same expression as in $4D$ case.\\

It is evident that the initial shear vanishes when $f_{2}=0,
(F_{2}+f_{3})=(F_{3}+f_{4})=......=0$ and hence even if the
initial shear is zero the dust distribution may be inhomogeneous
because the co-efficient $F_{2}$ may still be non-zero. This
makes in view of (26) $t_{s}$ a function of the comoving radial
co-ordinate $r$, so that the shell focusing singularity appears
at different $r$ at different instants. The nature of
singularities appearing in marginally bound cases $(f=0)$ is
therefore clearly distinct from the present case of
non-marginally bound case $(f\neq 0)$. This fact, however, has
not been taken care of by Joshi etal in their discussion in the 4
dimensional marginally bound space-time.\\

{\bf Acknowledgement:}\\

The authors are grateful to the referee for his valuable comments
and suggestions to improve the manuscript. They further thank the
members of the Relativity and Cosmology Research Centre,
Department of Physics, Jadavpur University for helpful
discussion. One of the authors (U.D) is thankful to CSIR (Govt. of
India) for awarding a Junior Research Fellowship.\\

{\bf References:}\\
\\
$[1]$  R. Penrose, {\it Riv. Nuovo Cimento} {\bf 1} 252 (1969);
in General Relativity, an Einstein
Centenary Volume, edited by S.W. Hawking and W. Israel (Cambridge Univ. Press, Cambridge, 1979).\\
$[2]$  P.S. Joshi, {\it Global Aspects in Gravitation and Cosmology} (Oxford Univ. Press, Oxford, 1993).\\
$[3]$  B. Waugh and K. Lake, {\it Phys. Rev. D} {\bf 38} 1315 (1988).\\
$[4]$  J. P. S. Lemos, {\it Phys. Lett. A} {\bf 158} 271 (1991);
{\it Phys. Rev. Lett.} {\bf 68} 1447 (1992).\\
$[5]$  P. S. Joshi and I. H. Dwivedi, {\it Class. Quantum Grav.}
{\bf 16} 41 (1999).\\
$[6]$  A. IIha and J. P. S. Lemos, {\it Phys. Rev. D} {\bf 55}
1788 (1997).\\
$[7]$  A. IIha, A. Kleber and J. P. S. Lemos, {\it J. Math.
Phys.} {\bf 40} 3509 (1999).\\
$[8]$  J. F. V. Rocha and A. Wang, {\it ibid} {\bf 17} 2589
(2000).\\
$[9]$  P.S. Joshi, N. Dadhich and R. Maartens, {\it Phys. Rev. D}
{\bf 65} 101501({\it R})(2002).\\
$[10]$  D. M. Eardley and L. Smarr, {\it Phys. Rev. D} {\bf 19}
2239 (1979); D. M. Eardley in Gravitation and Astrophysics, edited
by B. Carter and J. B. Hartle (NATO Advanced Study Institute,
Series B; Physics, Vol. 156) (Plenum Press, New York, 1986), pp.
229-235.\\
$[11]$  A. Benerjee, A. Sil and S. Chatterjee, {\it Astrophys.
J.} {\bf 422} 681 (1994); A. Sil and S. Chatterjee, {\it Gen.
Rel. Grav.} {\bf 26} 999 (1994).\\
$[12]$  S. Chatterjee, A. Banerjee and B. Bhui, {\it Phys. Lett. A}
{\bf 149} 91 (1990).\\
$[13]$  S. G. Ghosh and N. Dadhich, {\it Phys. Rev. D} {\bf 64}
047501 (2001).\\
$[14]$  S. G. Ghosh and A. Beesham, {\it Phys. Rev. D} {\bf 64}
124005 (2001).\\
$[15]$  S. G. Ghosh and A. Beesham, {\it Class. Quantum Grav.}
{\bf 17} 4959 (2000).\\
$[16]$  S. S. Deshingkar, P.S. Joshi and I. H. Dwivedi, {\it Phys.
Rev. D} {\bf 59} 044018 (1999).\\
$[17]$  T. Harada, H. Iguchi and K.I. Nakao, {\it Phys. Rev. D}
{\bf 61} 101502 (2000); {\it Prog. Theor. Phys.}
{\bf 107} 449 (2002) .\\

\end{document}